\documentclass[final,5p,times,twocolumn]{elsarticle}

\usepackage{subfigure}
\usepackage{graphicx}
\usepackage{epstopdf}
\usepackage{longtable}
\usepackage{amsmath}
\usepackage[load-configurations = abbreviations]{siunitx}
\usepackage{color}
\usepackage[usenames,dvipsnames,svgnames,table]{xcolor}
\usepackage{lineno,hyperref}
\usepackage{siunitx}

\modulolinenumbers[5]

\begin{document}

\author[SBAI,INFN-Roma1]{F.Mira}
\author[LNF]{M.Ferrario}
\author[INAF,INFN-Bo]{P.Londrillo}
\author[LNF]{A.Marocchino \corref{mycorrespondingauthor} }
\cortext[correspondingauthor]{Corresponding author}
\ead{alberto.marocchino@lnf.infn.it, alberto.marocchino@gmail.com}

\address[SBAI]{Dipartimento di Scienze di Base e Applicate per l'Ingegneria, Universit\'a degli studi di Roma "��La
Sapienza�", Via Antonio Scarpa, 14, 00161, Roma (Italy)}
\address[INFN-Roma1]{INFN sezione Roma1,Piazzale Aldo Moro, 2, 00185, Roma (Italy)}
\address[LNF]{Laboratori Nazionali di Frascati, INFN, Via E. Fermi 40, Frascati (Italy)}
\address[INAF]{Osservatorio Astronomico di Bologna, Via Piero Gobetti, 93/3, 40129, Bologna (Italy)}
\address[INFN-Bo]{INFN sezione Bologna,Bologna, Viale Berti Pichat, 6/2, 40127, Bologna (Italy)}

\title{Characterisation of beam driven ionisation injection in the blowout regime of Plasma Acceleration}

\date{\today}

\begin{abstract}
Beam driven ionisation injection is characterised for a variety of high-Z dopant. We discuss the region of extraction and why the position where electrons are captured influences the final quality of the internally-injected bunch. The beam driven ionisation injection relies on the capability to produce a high gradient fields at the bubble closure, with magnitudes high enough to ionise by tunnelling effect the still bounded electrons (of a high-Z dopant). The ionised electrons are captured by the nonlinear plasma wave at the accelerating and focusing wake phase leading to high-brightness trailing bunches. The high transformer ratio guarantees that the ionisation only occurs at the bubble closure. The quality of the ionisation-injected trailing bunches strongly and non-linearly depends on the properties of the dopant gas (density and initial ionisation state). We use the full 3D PIC code ${\tt ALaDyn}$ to consider the highly three-dimensional nature of the effect. By means of a systematic approach we have investigated the emittance and energy spread formation and the evolution for different dopant gases and configurations.
\end{abstract}

\maketitle

\section{Introduction}
Plasma acceleration is a promising novel accelerating technique to deliver high brightness, GeV energies beams with table-top size devices. Theoretical works \cite{TajimaDawson} \cite{Chen}, simulations \cite{Esarey} and experiments \cite{Leemans}\cite{Blumenfeld} \cite{FLAME} have demonstrated the possibility to inject and accelerate electron bunches in a plasma wave, driven either by a laser or a relativistic charged driver beam. However, the beam properties, such as emittance and energy spread, are still poor and novel schemes or approaches must be foreseen to retain bunch quality over the whole acceleration length.

\begin{figure}[htb]
\includegraphics[width=.4\textwidth]{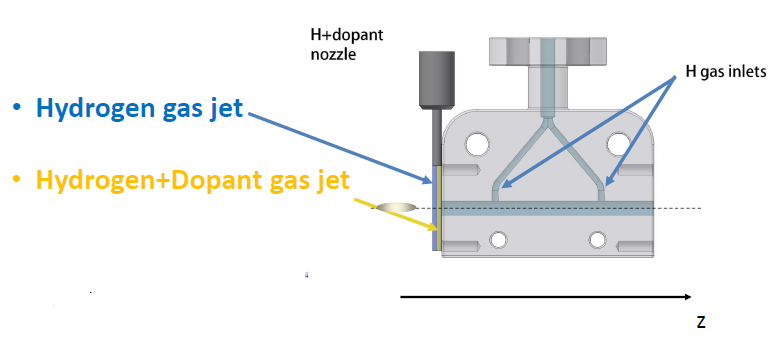}
\centering
\caption{Layout of plasma injector for Wake-Field Ionisation Injection. The driver beam propagates from left to right and encounters a two region gas jet (Hydrogen)+(Hydrogen+Dopant) as injection region, and a capillary for transport and acceleration.}   
\label{fig:capillary}
\end{figure}

A possible approach to retain the accelerated bunch quality is to use the \textit{external-injection} \cite{Hogan}. The external injection employs a conventional photo-injector technology to extract the bunch from a photo-injector and pre-accelerate it, while the plasma is used only as an extremely high gradient section. The power of this approach is based on the great reliability that photo-injector technology has reached \cite{Ferrario}. However, the requirement of a pre-accelerating structure determines a partial loss in terms of compactness. Conversely, magnetically assisted injection \cite{Vieira}, trojan horse injection \cite{Hidding}, ionisation injection \cite{Oz} \cite{Chen2006} \cite{mcguffey2010} \cite{Pak} \cite{clayton2010} \cite{pollock2011} and similarly derived approaches \cite{liu2011all} \cite{zeng2015multichromatic} \cite{mirzaie2015demonstration}, are based on the \textit{internal-injection} technique, i.e. the electrons from the plasma background are trapped in the accelerating phase induced by a driver. The pre-accelerating structure is no longer required but other technical problems arise (e.g. the synchronisation between the laser and the plasma wave, the witness beam degradation due to the driving beam betatron oscillations).
\begin{figure}[htb]
\includegraphics[width=0.5\textwidth]{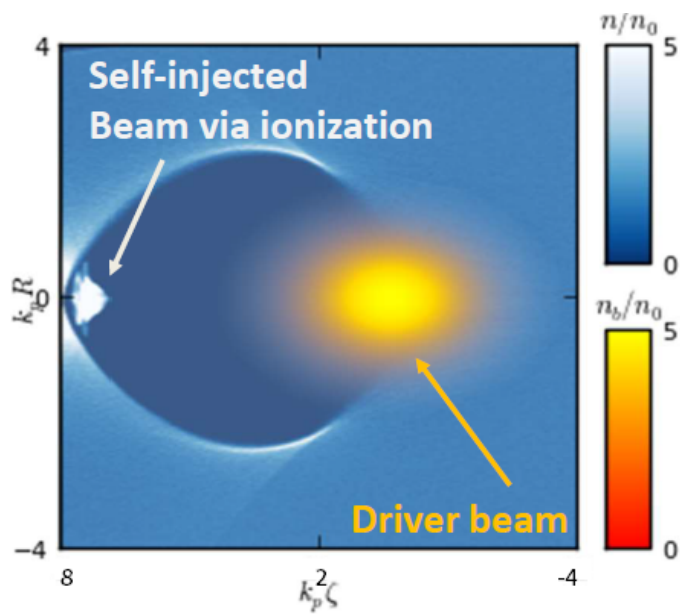}
\centering
\caption{Internal injected beam in the rear of the bubble via ionisation.}   
\label{fig:ioniz_inj_rho}
\end{figure}
Wakefield Ionisation Injection \cite{Ossa1} \cite{Ossa2} is a novel approach which employs the field developed in the beam driven blowout regime of plasma acceleration, in order to ionise a dopant element and to inject the stripped electrons directly in the accelerating and focusing phase of the plasma wave Fig \ref{fig:ioniz_inj_rho}.
In this scheme, the driving beam field contribute is neglected (assuming however its current is high enough to induce the bubble regime with high transformer ratios, \citep{massimo1}) via the choice of the dopant element with high ionisation potential, confined in a small region of a pre-ionised background plasma.\\
The high ionisation potential of the dopant gas ensures that it does not get ionised by the field of the driver, determining a much more stable injection since the ionising electric field only depend by the blowout configuration and is not affected by the driver betatron oscillation and depletion.
High current driver beams, several kA, are required to develop an intense wakefield able both to induce ionisation and to trap the ionised electrons in the accelerating and focusing phase of the plasma wave. Such beams can be either produced by a conventional photoinjector or a previous Laser driven injection stage in a gas jet device. The laser driver can be then both used as pre-ionisation source in the following witness injection device, as showed in Fig.\ref{fig:capillary}: \\ 

In this work we have characterised the internally-injected bunch quality varying the dopant gas. We chose three different high-Z dopant gases: Nitrogen, Argon and Neon. The systematic study has been conducted only for the injection region (a few hundreds of micron propagation distance) in the limit of a 10 kA - 1 GeV electron driver bunch. Simulations have been performed with the Particle in Cell (PIC) code ${\tt ALaDyn}$ \cite{ALaDyn1,ALaDyn2,ALaDyn3}, whose last code updates for Plasma Wakefield Acceleration (PWFA) is reported in \cite{Marocchino1}. We also recall that for quick estimation runs we can rely on the Architect hybrid approach \cite{Marocchino2,Marocchino3}. 

\section{Numerical setup}
The characterisation is operated by varying the dopant gas. We chose: Argon, Nitrogen, and Neon. The initial ionisation level of the dopant elements is chosen to match the pre-ionisation of the hydrogen, achieved via a low intensity laser beam or via an inductive discharge. Therefore, both the Argon and the Nitrogen were set as $Ar^{1+}$ and $N^{1+}$, since the first level ionisation potentials are 15.76 eV and 14.54 eV, close enough to the ionisation potential of the hydrogen (13.6 eV). This assumption does not apply to Neon where the first ionisation potential is 21.56~eV, and thus it is unlikely to be ionised by the pre-ionisation discharge/laser.

\begin{figure}[htb]
\includegraphics[width=.5\textwidth]{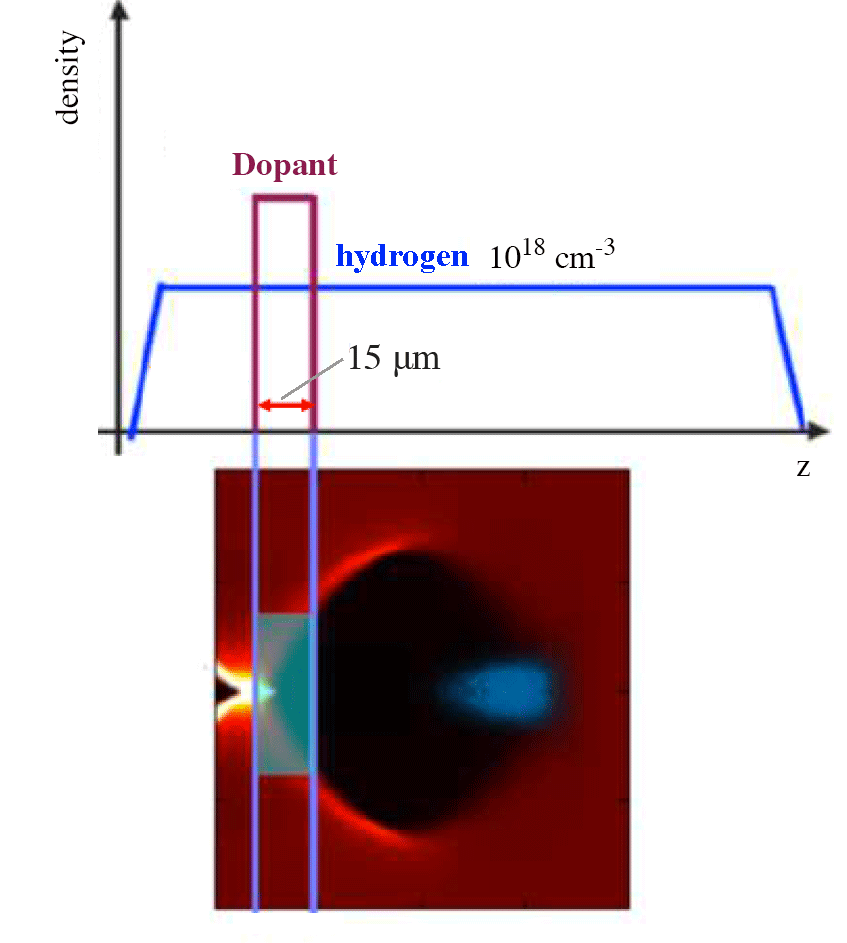}
\centering
\caption{Setup schematic, background and dopant profile schematic with reference to the bubble profile to highlight the region of electron ionisation and trapping.}   
\label{fig:numerical_experiment_setup}
\end{figure}

The background plasma density is $n_0 = 1.2\times10^{18}$ cm$^{-3}$, characterised by a wave-breaking field of $E_z \sim 110$~GV/m \cite{Dawson}. The dopant percentage, with respect of the background density, was chosen to internally-inject a 10~pC trailing bunch. The dopant layer is 15~$\mu$m thick, as shown in Fig.~\ref{fig:numerical_experiment_setup}; such length was chosen in order to reduce as much as possible the slice energy spread avoiding injection of dark current.

The driver is a 1~GeV, 10~kA current electron beam, with $\sigma_r = 4$~$\mu$m and $\sigma_z = 7$~$\mu$m. The bunch is evolved for 600~$\mu$m, a distance that allows to reach an asymptotic behaviour of the internal-injected beam characteristics, specifically, emittance and energy spread. We focus our attention only to the injection phase, reducing as much as possible computational costs. The simulation box is set with 416 cells along the longitudinal coordinate $z$, and 520$\times$520 cells along the transverse plane ($x$-$y$). The computational cell size is $\Delta_z = 0.2$~$\mu$m longitudinally and $\Delta_x = \Delta_y = 0.08$~$\mu$m transversely. Each cell contains 8 macro-particles for the electrons, and 1 ion macroparticle (only in the region where we account for the dopant layer).

\section{Injection Volume}
\begin{figure}[htb]
\includegraphics[width=.5\textwidth]{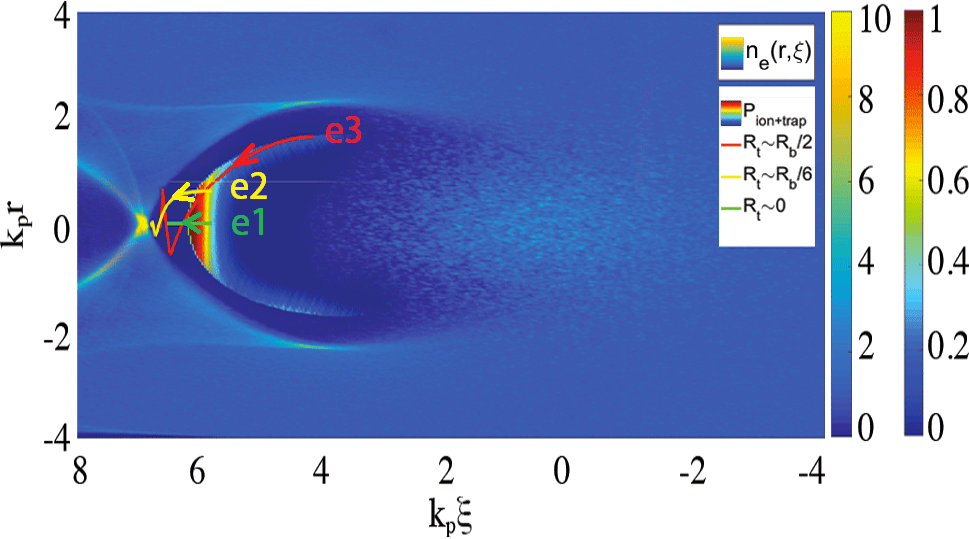}
\centering
\caption{Ionisation probability over-imposed to the density colormap, with the trajectory of three electrons for the $Ar^{1+}$ case. The ionisation rate is plotted with a jet colormap from 0 to 1, on the background the density colormap from 0 saturated at 10 (yellow color). The Figure also reports the trajectory of three electrons $e1$ (green, on axis), $e2$ (yellow) and $e3$ (red).}   
\label{fig:trajectories}
\end{figure}

Ionised electrons are approximately at rest when they are stripped from the dopant element ion, their thermal motion is negligible with respect to the collective motion that sustains the plasma wave. The trapping condition \cite{Mora}\cite{Pak}, written as function of generalised potentials, $\phi$ and $\textbf{A}$, is:
\begin{equation}\label{eq:trapping}
\psi(\xi_f,r_f)-\psi(\xi_i,r_i)=-\frac{mc^2}{e}\left(1-\frac{1}{\gamma_w}\right)
\end{equation}
where $\psi(\xi,r)=\phi(\xi,r)-v_\phi A_z(\xi,r)$ is the wake potential for a plasma wave propagating at $v_\phi\sim c$, and $\gamma_w$ is its Lorentz factor, $\textbf{A}$ the vector potential and $\xi$ the co-moving coordinate. In PWFA scenarios the approximations $v_\phi\rightarrow c$ and $\gamma_w\rightarrow\infty$ hold, greatly simplifying the theoretical trapping condition to the simple expression $\Delta\psi=-1$. We have used the normalised wake potential $\psi e/mc^2$. The tunnel ionisation process for high-Z atoms can be described by the ADK ionisation formulation \cite{ADK}, which relates the ionisation probability to the ion ionisation potential and to the magnitude of the electric field directly acting on the ion of interest. The ionisation probability $P_{ionz}=(\xi_i,r_i)$ therefore depends on the blowout geometrical configuration, since the electric field implied in ionisation roughly scales linearly as the bubble radius $E_z\sim R_b$. The bubble radius $R_b$ in turn depends on the driving beam normalised current \cite{Lu} $R_b=2.5\sqrt{\Lambda}=2.5\sqrt{I_b/I_A}$, where $I_A=17 kA$ is the Alfv\`{e}n current. Besides, the wake potential is related to the accelerating field by $\partial_z\psi=-E_z$ \cite{Lu}, and therefore scales as $\psi\sim R_b^2$.
By combining the ionisation probability and the trapping threshold Eq.~(\ref{eq:trapping}), it is possible to define an ionisation volume inside the blowout region which satisfies the trapping condition for any $(r,\xi)$, which we will call \textit{injection volume}. In Fig. \ref{fig:trajectories} we show, as an example, the injection volume for the dopant element $Ar^{1+}$, the colours corresponding to the ionisation probability.\\
The injection volume for $N^{1+}$ is very similar, both in terms of local ionisation probability and injection volume, while the local ionisation probability for $Ne^0$ is much lower and does not reach unity in the injection volume. This difference compared to $Ar^{1+}$ and $N^{1+}$ leads to a lower trapped charge along the injection process.
The highest ionisation probability, i.e. higher than $0.5$, is confined in a hemispherical area of radius $\sim R_b/3$ and length $\sim R_b/6$, where $k_p$ is the plasma wavenumber. The electron population ionised in this central region is therefore inherently characterised by a very low divergence, since the transverse momentum for electrons born at rest in the blowout region only depends on the transverse force $W_\perp$ exerted by the positive ion column \cite{Lu} scaling as $W_\perp\sim r$. The trajectory for two electrons born in these locations is reported in Fig.~\ref{fig:trajectories} and labeled as $e1$ and $e2$. On the other hand, the light blue shadow, similar to a hollow cylinder having $R_t\sim R_b/2$ and length $\Delta z\sim R_b$, is responsible for the higher divergence electron population, as $e3$, since electrons ionised in that region feel a much higher transverse force. However, this region corresponds to a lower ionisation probability area, and therefore influences the whole beam divergence less than the previous central region, representing thus only the tails of the self-injected beam distribution function.\\
The energy spread of the internally-injected beam is related to the geometry of the injection volume as well. If we do not consider beam loading effects, the energy distribution of the internally-injected beam is directly related to the beam length, which depends on the injection volume radius. During the trapping process the electrons on axis, (see trajectory $e1$), are subject only to a longitudinal force and therefore reach the phase velocity of the plasma wave $v_\phi\rightarrow c$ faster than electrons off axis (see trajectories $e2$ and $e3$), which have residual transverse momentum that must damp completely before their longitudinal velocity matches the speed of light. Once the collection of ionised electrons is over, the difference in final longitudinal positions between the on axis and off axis electron populations, i.e. the internal-injected beam length, implies different magnitudes of accelerating field along the longitudinal coordinate $\xi$, and therefore the generation of correlated energy spread. 
\section{Analysis for different dopant elements}
\begin{figure}[htb]
\includegraphics[width=0.45\textwidth]{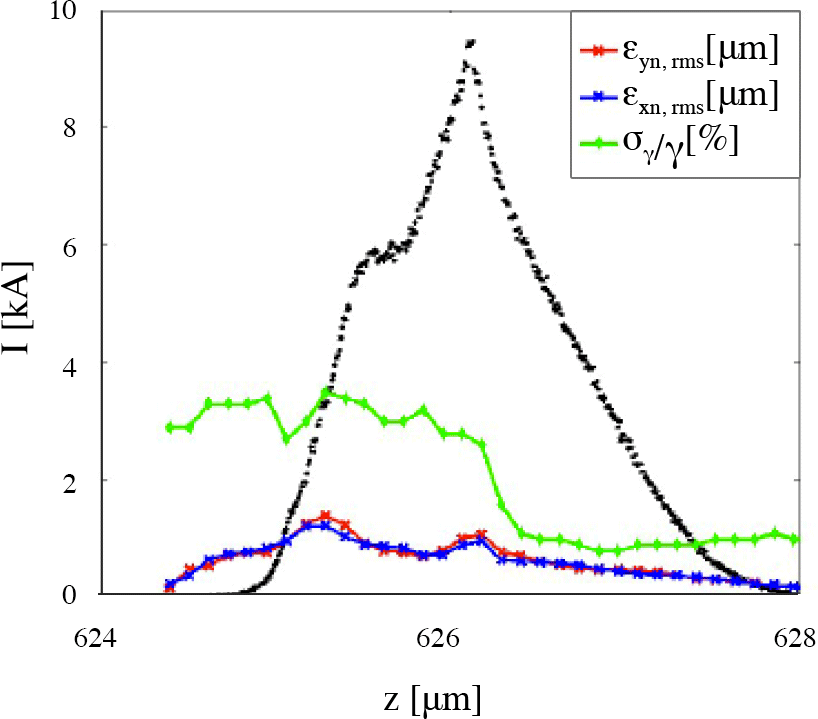}
\centering
\caption{Slice analysis for the Nitrogen case at 600~$\mu$m. The beam current profile is superimposed on the relative energy spread (green curve) and the $\varepsilon_{xn,rms}$  and $\varepsilon_{xn,rms}$. The slice dimension is 0.1~$\mu$m.}   
\label{fig:slice_analysis}
\end{figure}

\begin{figure}[htb]
\includegraphics[width=.5\textwidth]{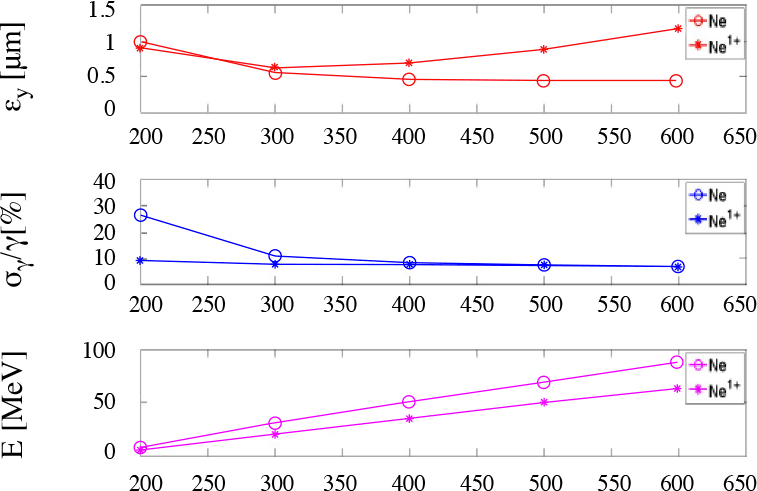}
\centering
\caption{Initial ionisation level influence on beam properties: Ne ground state and Ne$^{1+}$ comparison over emittance (red curve), relative energy spread (blue curve) and energy gain (purple curve).}   
\label{fig:Neon_comparison}
\end{figure}

In this section we describe the results obtained via PIC simulations varying the dopant element used as active electron source.
\begin{table}[htb]\centering \caption{internal-injected bunch properties at 600~$\mu$m \label{tab:simulation_results_1}}
\begin{tabular}{l c c c c }\hline\hline
\multicolumn{1}{c}{\textbf{Species}} &\textbf{Q [\rm{pC}]}&\textbf{$I_{rms}$ [\rm{kA}]}& \textbf{$\sigma_r [\rm{\mu m}]$}& \textbf{$\sigma_z[\rm{\mu m}]$}\\ \hline
$N^{1+}$ & 22.2 & 4.7 & 0.69 & 0.57\\
$Ar^{1+}$ & 26.7 & 4.49 & 0.92 & 0.71\\
$Ne^{0}$ & 22.5 & 4.04 & 0.74 & 0.67\\
\hline\end{tabular}
\end{table} 
\begin{table}[htb]\centering \caption{internal-injected bunch quality properties at 600~$\mu$m \label{tab:simulation_results_2}}
\begin{tabular}{l c c c }\hline\hline
\multicolumn{1}{c}\textbf{$\frac{\sigma_\gamma}{\gamma} [\%]$}& \textbf{$\varepsilon_{n,rms}[\rm{\mu m}]$}& \textbf{$\sigma_\gamma [\rm{MeV/mc^2}]$}&\textbf{$\gamma [\rm{MeV/mc^2}]$}\\ \hline
\;4.7 & 0.96 & 3.01 & 64.3\\
\;5.9 & 0.96 & 3.58 & 60\\
\;7 & 1.15 & 4.5 & 64.2\\
\hline\end{tabular}
\end{table}\\
In Table~\ref{tab:simulation_results_1} and Table~\ref{tab:simulation_results_2} we report on the internally-injected beam properties at 600~$\mu$m. We observe that beam parameters are similar, since the ionisation volume from where they originate is similar, similarity that is reflected on bunch quality. We conclude that for high Z dopant the choice of the material can be operated with some freedom with no major influence on the bunch quality.
\begin{figure}[htb]
\includegraphics[width=0.5\textwidth]{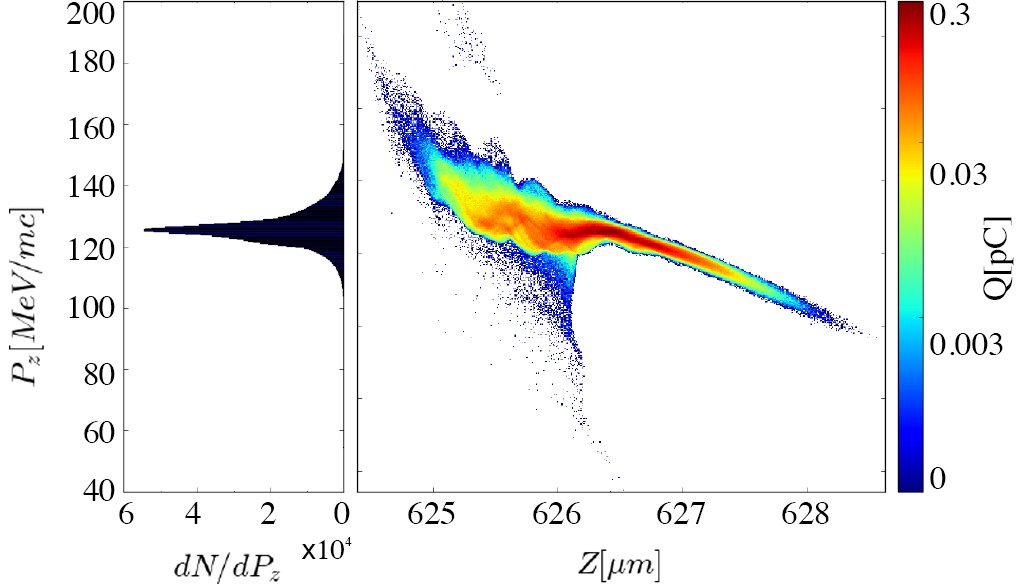}
\centering
\caption{Longitudinal phase-spaces and energy spectra of Nitrogen for the injection distance of $600\;\rm {\mu m}$. Beam loading effects due to the high charge injected, induce low energy spread.}   
\label{fig:zPz_hist_06_N}
\end{figure}
The internally-injected beam longitudinal phase space presented in Fig.\ref{fig:zPz_hist_06_N} corresponds to the $N^{+1}$ dopant case. The charge is mainly concentrated in the FWHM, around 75 $\%$ of the total charge, denoting a well confined and peaked profile. If we focus our attention in this region, we appreciate how the energy spread and the emittance have particularly low values, i.e. $<3\%$ and $<$1$\mu$m respectively.
A positive beam-loading effect explains the low energy spread. The region of the bunch within the FWHM feels an almost flat accelerating field along the plasma acceleration and transport, and therefore does not experience absolute energy spread growth. On the other hand, the distribution tails at higher and lower energy feel a large difference in accelerating field, but their contribution to the behaviour of the whole beam is smaller, since they contain only $\sim25\%$ of the beam charge. The slice analysis of Fig.\ref{fig:slice_analysis} reports how the betatron decoherence, due to different ionisation phases along the dopant layer, corresponds to the high energy tail of the beam distribution, located in the highly non-linear area at the back of the bubble.\\ Of more practical use are the data reported in Fig.~\ref{fig:Neon_comparison} where the the beam injection and transport integrated quantities for the pre-ionised Neon dopant case $Ne^{1+}$ and the ground state Neon dopant case $Ne^0$ are reported.
It is clear from this comparison how the ionisation volume determines different beam properties such as emittance and energy spread.  Since the ionisation potential for $Ne^0$ is much lower than for $Ne^{+1}$, respectively 21.56 eV and 40.96 eV, the ionisation volume in the first case is much wider. This leads to more than double of emittance for $Ne^{0}$ compared to $Ne^{+1}$, i.e. 0.45 $\rm{\mu m}$ vs 1.15 $\rm{\mu m}$.
The same argument applies to the energy spread. In the pre-ionsed case $Ne^{1+}$ the injected charge is lower than the ground state case, respectively 0.1\; pC vs $\sim$20\; pC, and the beam duration is smaller as well, i.e. $\sigma_{z,Ne^0}\sim 0.2\; \rm{\mu m}$ vs $\sigma_{z,Ne^{1+}}\sim 0.67\; \rm{\mu m}$. No beam loading effect occurs, and a moderate energy spread is induced because of the short length. The same value of $\sigma_\gamma/\gamma\sim7\%$ is induced in the ground state case $Ne^0$, now because even though the beam is longer, the higher charge strongly loads the field, flattening the profile along the beam longitudinal dimension. The beam loading effect is even clearer if we compare the energy gain for the two cases, 87 MeV for $Ne^{0}$ and 64 MeV for $Ne^{1+}$, where for the lower injected charge for the pre-ionised case determines a less loaded and thus a more intense accelerating field felt along the transport.\\
\section{Conclusions}
In this work we presented the results obtained via PIC simulations in different scenarios of the beam driven Wakefield Ionisation Injection scheme. Low energy spread, low emittance, and high current, and thus high brigthness internally-injected beams, are inherently generated in every dopant element configuration, since the core of the injection volume for the different dopant elements does not differ significantly. The self-consistent simulations, and the single particle dynamics show how emittance and energy spectrum of the internal-injected beam depend on the injection volume geometry, and therefore on the blowout configuration, the dopant element Z, and its initial ionisation state.

\section{Acknowledgements}
The authors would like to acknowledge INFN-CNAF for providing the computational resources and support required for this work. A. M. and F. M. also acknowledge the CINECA award under the ISCRA initiative, for the availability of high performance computing resources and support. This work was supported by the European Union Horizon 2020 research and innovation programme under grant agreement N. 653782. The authors would like to acknowledge F. Massimo for useful discussions.


%
%
%
%

\end{document}